\documentclass[12pt]{article}
\usepackage{blindtext} 
\usepackage{amsmath}

\usepackage[english]{babel} 

\usepackage[hmarginratio=1:1,top=32mm,columnsep=20pt]{geometry} 
\usepackage[hang, small,labelfont=bf,up,textfont=it,up]{caption} 
\usepackage{booktabs} 
\usepackage{enumitem} 
\setlist[itemize]{noitemsep} 

\usepackage{abstract} 

\usepackage{titlesec} 
\renewcommand\thesection{\Roman{section}} 
\renewcommand\thesubsection{\roman{subsection}} 
\titleformat{\section}[block]{\large\scshape\centering}{\thesection.}{1em}{} 
\titleformat{\subsection}[block]{\large}{\thesubsection.}{1em}{} 

\usepackage{fancyhdr} 
\usepackage{titling} 

\usepackage{hyperref} 

\pretitle{\begin{center}\Huge\bfseries} 
\posttitle{\end{center}} 
\title{Massive open string and totally-noncommutativity in space and momentum coordinates trough Fadeev Jackiw approach} 
\author{%
\textsc{N. Mansour$^1$, E. Diaf$^2$ and M.B. Sedra$^1$ }   \\  
\normalsize $^1$ SIMO-Lab, Faculty of Sciences, Ibn Tofail University, Kenitra, Morocco \\ 
\normalsize $^2$ Department of Physics, OLMAN-RL Pluridisciplinary Faculty of Nador, \\
\normalsize University Mohammed First, Oujda B.P N 300, Selouane, 27000 Nador, Morocco. \\
\normalsize \href{mailto:nmansour@live.fr}{nmansour@live.fr} \normalsize \href{mailto:eldiaf@gmail.com}{eldiaf@gmail.com}\normalsize \href{mailto:mysedra@yahoo.fr}{mysedra@yahoo.fr}}
\date{} 
\renewcommand{\maketitlehookd}{%
\begin{abstract}
In this work we will show that it is possible to construct totally noncommutativity in the cas of massive bosonic string in the D-brane background with a non-vanishing constant B-field, by extending
the well known Faddeev-Jackiw formalism proposed by L.D. Faddeev and R.
Jackiw  [18].\\~~\\
\textbf{Key words}: Massive bosonic  string, F.J. quantization , Noncommutative geometry.
\\~~\\
\textbf{PACS} numbers : 11.25.Yb; 11.25.-w; 11.25.Hf; 02.40.Gh; 11.25.Uv
\end{abstract}
}

\begin{document}

\maketitle

\section{Introduction}
In recent years, there is an increasing interest in the studying of noncommutative (NC) geometry  [1] and its application to physical problems. It was shown that for an open strings attached to a D-branes in the presence of background B-field would induce noncommutativity of space  in its end points
[2, 3, 4, 5, 6, 7, 8, 9, 10, 11, 12, 13, 14, 15, 16, 17, 19]. 
The primary aim of this paper is to study  massive bosonic strings propagating in the presence of a background two form field $ B_{\mu\nu} $, this study has become important because it exhibits a manifest of totally noncommutative structure among the momentum and space coordinates, somme  approaches have been taken to obtain such results [17], in the present paprer we employ the Faddeev-Jackiw symplectic formalism to study the massive bosonic string in externel antisymmetric B-field to build up the totally noncommutativity wich include the noncommutativity of momentum coordinates which  is the central perpos of this paper.\\
This paper is organized as follows. In the first section we introduce the model of the action of massive bosonic string in  external antisymmetric B-field in its discrete version and the Faddeev-Jackiw methodology employed to this model to build up the totally non commutative structure of space and momentum we think this method is more easy to get this complicate result.In the last section we give our conclusion.
\section{The Faddeev-Jackiw method}
In this section we discuss the methodology of F.J. approach to quantize the singular systems. In this formalism, we first write the Lagrangian of a singular system into the first-order form as follows:\\
\begin{align}
L^0=a^0_k(\xi)\partial_\tau \xi^k-V(\xi)
\end{align}
Where  $ \xi^i $ is called the symplectic variables, $V (\xi)$ is called the symplectic potential. The first-order form can be implemented by introducing some auxiliary variables $(a_k)$ such as the canonical momentum.
The Euler-Lagrange equations of motion for Lagrangian (1) can be written as
\begin{align}
f^0_{ij}(\xi)\dot{\xi}^j=\frac{\partial V(\xi)}{\partial \xi^i} ~~~~~~~~ ( i=1,2,3,...,n),
\end{align}
where $f_{ij}$ is so-called symplectic matrix with following explicit form:
\begin{align}
f^0_{ij}=\frac{\partial a^0_j}{\partial \xi^i}-\frac{\partial a^0_i}{\partial \xi^j}
\end{align}
If it is non singular we define the commutators of the quantum theory (if there is no ordering
problem for the corresponding quantum operators) as:
\begin{align}
[A(\xi),B(\xi)]=\frac{\partial A}{\partial \xi_i}(f^0)^{-1}_{ij} \frac{\partial B}{\partial \xi_j}
\end{align}
If the matrix $ (3) $ is singular we find the zero modes that satisfy $ f^0_{ij}v^\alpha_j = 0$ and the corresponding constraints:
\begin{align}
\Omega^\alpha=v^\alpha_l \frac{\partial V}{\partial \xi_l  }=0
\end{align}
Now, we modify our original Lagrangian by introducing the constraint term multiplied with some Lagrange multipliers $\lambda^\alpha $ as
\begin{align}
L^1=a^0_k(\xi)\dot{\xi}_k+\dot{\lambda^\alpha} \Omega^\alpha-V(\xi)=a^1_r(\tilde{\xi})\dot{\tilde{\xi}}-V(\xi)
\end{align}
where we introduced the new notation for the extended variables: $\tilde{\xi}^r=(\xi^k,\lambda^\alpha) $We find now
the new matrix $f^1_{ij} $
\begin{align}
f^1_{ij}=\frac{\partial a^1_j}{\partial \tilde{\xi}^i} -\frac{\partial a^1_i}{\partial \tilde{\xi}^j} 
\end{align}
If $ f^1$ is not singular we define the quantum commutators as
\begin{align}
[A(\tilde{\xi}),B(\tilde{\xi})]=\frac{\partial A}{\partial \tilde{\xi}_i}(f^1)^{-1}_{ij} \frac{\partial B}{\partial \tilde{\xi}_j}
\end{align}
This process of incorporating the constraints in the Lagrangian is repeated until a non singular
matrix is found.\\
\section{Massive bosonic string in antisymmetric B-filed and totally noncommutativity}
A string is a 1-dimensional object moving in D-dimensional space-time (1 time, D-1 space dimensions). As the string moves in time it sweeps a 2-dimensional area in D-dimensional space-time, this called the worldsheet of the string denoted  $ \Sigma $. This string parameterized by a parameter $ \sigma, 0<\sigma<l $ (for somme finit $ l $). As the string moves in time, every point on the string, describes a trajectory in space-time, this trajectory can be parameterized by a variable $ \tau $. $ \tau $ could have the range $ -\infty <\tau < +\infty $. Every point on the worldsheet is parameterized by the pair $ \sigma^\alpha=(\tau,\sigma)$. $ g_{\alpha\beta}=diag(-,+) $, $\epsilon^{01}=-\epsilon^{01}=1 $ is the anti-symmetric symbol in two dimensions, $ B_{\mu\nu}=-B_{\nu\mu} $,  $ \eta_{\mu\nu}=diag(-,+,...,+) $ and the length of the string is $ \pi $.\\
 The space-tame position of a point $ (\tau, \sigma) $ on the string worldsheet $ \Sigma  $ is given by the fuction $  X^\mu (\tau,\sigma), (\mu= 0,1,...D-1)$ \\

The action for massive bosonic open string with its endpoints attached on a D-brane in the presence of an NS B-field is :\\
\begin{align}
S=&-\frac{1}{4\pi\alpha'}\int_\Sigma d\xi^2 (g_{ij} \eta^{\alpha\beta}\partial_\alpha X^i \partial_\beta X^j +g_{ij} m^2 X^iX^j+ \epsilon^{\alpha\beta} B_{ij}\partial_\alpha X^i \partial_\beta X^j)\\
\nonumber
=&-\frac{1}{4\pi\alpha'}\int_\Sigma d\xi^2[(\dot{X}^i)^2-(X'^{i})^{2}+m^2(X^i)^{2}-2B_{ij}\dot{X}^i X'^j]
\end{align}
Where "dot" and "prime" represent differentiation with respect to $\tau $ and $ \sigma $ respectively.\\
The massless cas $(m=0)$ is studied in different aspects by several authors [10, 12, 13, 14] with the well-known result of noncommutativity of space coordinates of the end points.\\
We will consider a discret version of the string in which we replace the continuous coordinate $ \sigma $ with range $(0,\pi)  $ by a discrete set coresponding to intervals of length $ \epsilon $. Representing the string coordinates at the end points of the N intervals as $  X_0^\mu,X_1^\mu,...X_{N-1}^\mu, X_N^\mu  $, and the resulting discretized lagrangian  in an external B-field associated with the action gived by :
\begin{align}
\nonumber
 S=&\frac{1}{4\pi\alpha'}\int dt[-\epsilon\eta_{\mu\nu}\dot{X}^\mu_0\dot{X}^\nu_0-\epsilon\eta_{\mu\nu}\dot{X}^\mu_1\dot{X}^\nu_1-...-\epsilon\eta_{\mu\nu}\dot{X}^\mu_{(N-1)}\dot{X}^\nu_{(N-1)}-\epsilon\eta_{\mu\nu}\dot{X}^\mu_N\dot{X}^\nu_N\\
 \nonumber 
&+ \frac{1}{\epsilon}\eta_{\mu\nu}(X_1-X_0)^\mu(X_1-X_0)^\nu+\frac{1}{\epsilon}\eta_{\mu\nu}(X_2-X_1)^\mu(X_2-X_1)^\nu+...\\
\nonumber
&\frac{1}{\epsilon}\eta_{\mu\nu}(X_{N-1}-X_{N-2})^\mu(X_{N-1}-X_{N-2})^\nu+\frac{1}{\epsilon}\eta_{\mu\nu}(X_N-X_{N-1})^\mu(X_N-X_{N-1})^\nu\\
\nonumber
&+\eta_{\mu\nu}m ^2\epsilon X^\mu_0X^\nu_0+\eta_{\mu\nu}m ^2\epsilon X^\mu_1X^\nu_1+...+\eta_{\mu\nu}m ^2\epsilon X^\mu_{N-1}X^\nu_{N-1}+\eta_{\mu\nu}m ^2\epsilon X^\mu_NX^\nu_N\\
\nonumber
&2B_{\mu\nu}\dot{X}^\mu_0(X_1-X_0)^\nu+2B_{\mu\nu}\dot{X}^\mu_1(X_2-X_1)^\nu+...+2B_{\mu\nu}\dot{X}^\mu_{N-1}(X_{N-1}-X_{N-2})^\nu\\
& +2B_{\mu\nu}\dot{X}^\mu_N(X_N-X_{N-1})^\nu]
\end{align}
Where we take equal spacing $ \int d\sigma=\epsilon \Sigma_n $. The continuum limit is achieved by the following replacements:
\begin{align}
\nonumber
&N\rightarrow \infty\\
\nonumber
&\epsilon \rightarrow 0\\
\nonumber
&N \epsilon \rightarrow l\\
&n\epsilon \rightarrow \sigma\\
\nonumber
&X_n \rightarrow X(\sigma,\tau)\\
\nonumber
&\frac{X_{n+1}-X_n}{\epsilon}\rightarrow\partial_\sigma X(\sigma,\tau)
\end{align}
Using the Euler Lagrange equations $\frac{d}{d\tau}(\frac{\partial L}{\partial \dot{X^\mu}})-\frac{\partial L}{\partial X^\mu}=0 $, we get the equation of motion for respectively $ X^\mu_0 , X^\mu_i , X^\mu_N $ with $ 1\leq n \leq N-1 $ .
\begin{align}
&\epsilon \partial_0^2 X^\mu_0=\frac{1}{\epsilon}(X_1-X_0)^\mu+\epsilon m^2 X_0^\mu\\
&\epsilon \partial_0^2 X^\mu_i=\frac{1}{\epsilon}(X_{(i+1)}-2 X_i+X_{(i-1)})^\mu+\epsilon m^2 X^\mu_i, i\neq 0,N.\\
&\epsilon \partial_0^2 X^\mu_N=\frac{1}{\epsilon}(X_N-X_{(N-1)})^\mu+\epsilon m^2 X_N^\mu.
\end{align}
When we take the limit $ \epsilon\rightarrow 0 $  and consider that $X^i_1\rightarrow X^i_0 $ and $X^i_{N-1}\rightarrow X^i_N $ the equations
for $X^i_0 $ and $X^i_N $ give the open string boundary conditions.\\
The BCs in the discrete version are:
\begin{align}
&\frac{1}{\epsilon}(X_1-X_0)^\mu -B^\mu_{~~~\nu}\dot{X}^\nu_0=0\\
&\frac{1}{\epsilon}(X_N-X_{N-1})^\mu -B^\mu_{~~~\nu}\dot{X}^\nu_N=0
\end{align}
The equations for points $X^i_n $ give no contribution at order zero in $\epsilon $ but to order one in $\epsilon $ (dividing by $\epsilon $ and then taking the limit $ \epsilon\rightarrow 0 $) they take the form of the standard equation of
motion for the string coordinates :\\
\begin{align}
\partial^2_\tau X^i_n - \partial^2_\sigma X^i_n-m^2 X^i_n=0
\end{align} 
for the above discrete version of the BCs, we see that it is sufficient for us only to take two points , say $ X^{\mu}_0,X^{\mu}_1  $   into consideration, in other words  we can consider just the end-points $ \sigma = 0 $, and for the other endpoint 
$ \sigma = 1 $ we will get the same result with opposite signs (open strings are noncommutating only at the ends, where open strings are attached to the brane, i.e.$ \sigma = \sigma_0 = 0 ; \sigma = \sigma_0 = \pi $ , and the
center of mass coordinates are commuting.)
 The action for these two points in the discrete form are :
\begin{align}
\nonumber
S=&\frac{1}{4\pi\alpha'}\int dt[-\epsilon\eta_{\mu\nu}\dot{X}^\mu_0\dot{X}^\nu_0-\epsilon\eta_{\mu\nu}\dot{X}^\mu_1\dot{X}^\nu_1+\frac{1}{\epsilon}\eta_{\mu\nu}(X_1-X_0)^\mu(X_1-X_0)^\nu\\
\nonumber
&+\frac{1}{\epsilon}\eta_{\mu\nu}(X_2-X_1)^\mu(X_2-X_1)^\nu+\eta_{\mu\nu}m ^2\epsilon X^\mu_0X^\nu_0+\eta_{\mu\nu}m ^2\epsilon X^\mu_1X^\nu_1\\
&+2B_{\mu\nu}\dot{X}^\mu_0(X_1-X_0)^\nu+2B_{\mu\nu}\dot{X}^\mu_1(X_2-X_1)^\nu]
\end{align}
According to Faddeev Jackiw. We shall re-express the action (18) in a first-order form. In doing so , we take the first term $ -\epsilon\eta_{\mu\nu}\dot{X}^\mu_0\dot{X}^\nu_0 $ of the action and we solve the BCs (15) into it, we get the  result as following
\begin{align}
&-\epsilon\eta_{\mu\nu}\dot{X}^\mu_0\dot{X}^\nu_0+\frac{1}{\epsilon}\eta_{\mu\nu}(X_1-X_0)^\mu(X_1-X_0)^\nu+2B_{\mu\nu}\dot{X}^\mu_0(X_1-X_0)^\nu\\
\nonumber
&=(B^{-1} M)_{\mu\nu}(X_1-X_0)^\mu \dot{X}^\nu_0.
\end{align}
Where $ M=1-B^2 $.\\~~~~\\
We obtain the action which is related with the $ X^0_\mu $ written as the first-Order form
\begin{align}
S=&\frac{1}{4\pi\alpha'}\int dt[(B^{-1}M)_{\mu\nu}(X_1-X_0)^\mu\dot{X}^\nu_0+L_{x_1}]
\end{align}
Where \\
\begin{align}
\nonumber
L_{x_1}=&-\epsilon \dot{X}_{1\mu}\dot{X}_1^\mu +2B_{\mu\nu}\dot{X}^\mu_1(X_2-X_1)^\nu+\frac{1}{\epsilon}(X_2-X_1)_\mu(X_2-X_1)^\mu\\
 &+\eta_{\mu\nu}m ^2\epsilon X^\mu_0X^\nu_0+\eta_{\mu\nu}m ^2\epsilon X^\mu_1X^\nu_1
\end{align}
To rewrite the second term $ -\epsilon\eta_{\mu\nu}\dot{X}^\mu_1\dot{X}^\nu_1 $ of the action we proceed as following :\\~~\\
As there are no constraints on the variables $ X_{1\mu} $, the Lagrangian $L_{x_1} $ is treated in the standard way, that is we introduce the conjugate momenta $ (P_{1\mu}) $ to the point $ X^\mu_1 $ It is defined as
\begin{align}
P_{1\mu}=\frac{\delta S}{\delta \dot{X}^\mu_1}=B_{\mu\nu}(X_2-X_1)^\nu-\epsilon\dot{X}_{1\mu}.
\end{align}
And the hamiltonian corresponding to $ L_{x_1}$ is 
\begin{align}
\nonumber
H_{x_1}&=P_{1\mu}\dot{X}^\mu_1-L_{x_1}\\
\nonumber
&=-\frac{1}{2\epsilon} P_{1\mu} P^\mu_1-\frac{1}{2\epsilon} M_{\mu\nu}
(X_2-X_1)^\mu (X_2-X_1)^\nu\\
&+\frac{1}{\epsilon}P^\mu_1 B_{\mu\nu}(X_2-X_1)\nu+\eta_{\mu\nu}m ^2\epsilon X^\mu_0X^\nu_0+\eta_{\mu\nu}m ^2\epsilon X^\mu_1X^\nu_1.
\end{align}
the following coupled Differential  equations in there discret form are used as  constraints and we suppose that are valid throught all the string 
\begin{align}
&B_{\mu\nu} P^\nu_1+ M_{\mu\nu} (X_2-X_1)^\nu=0\\
&P_1^\nu-P_0^\nu=m^2 B_{\mu\nu} X_0^\nu
\end{align}
We introduce (22),(24) and (25) in the lagrangian and we neglect the term in $B^2 $, then the action which is related with the $  X^\mu_0 $ and $  X^\mu_1 $ is translated the  into the first order form,
\begin{align}
\nonumber
S=&\frac{1}{4\pi\alpha'}\int dt[(B^{-1}M)_{\mu\nu}(X_2-X_0)^\mu\dot{X}^\nu_0-P_{1\mu} \dot{X}_0^\mu+P_{0\mu} \dot{X}^\mu_1+m^2 B_{\mu\nu} X^\mu_1\dot{X}^\nu_1\\
\nonumber
&~~~~~~~~~~~~~~-H_{x1}]
\end{align}
A set of symplectic variables , $ \xi^\mu_i=(X^\mu_0,X^\mu_1,P^\mu_0) $, and the corresponding canonical one-form 
$ a_{i\mu}=((B^{-1}M)_{\mu\nu}(X_2-X_0)^\nu+P_{1\mu};P_{0\mu}+m^2 B_{\mu\nu} X^\mu;0) $ can be read from action (12).We restrict ourselves to the D2-brane, in this case $ \mu,\nu=1,2 $ so we have this symplectic variables:
\begin{align}
\xi^\mu_i= \lbrace X^1_0,X^2_0,X^1_1,X^2_1,P^1_0,P^2_0\rbrace
\end{align}
and the corresponding canonical one-form:

\begin{align}
\nonumber
 a_{i\mu}=&  \lbrace(B^{-1}M)_{\nu 1}(X_2-X_0)^\nu+P_{11};(B^{-1}M)_{\nu 1}(X_2-X_0)^\nu+P_{12};\\
 &P_{0 1}+m^2 B_{\mu 1}X^\mu_1;P_{02}+m^2 B_{\mu 2}X^\mu_2;0;0\rbrace
\end{align}
These result in the symplectic two-form matrix $ f $

\begin{align}
(g_{\mu\nu})_{ij}=\frac{\partial(a_\nu)_j}{\partial(\xi^\mu)^i}-\frac{\partial(a_\mu)_i}{\partial(\xi^\nu)^j}
\end{align}
$f $ is a $ 6\times 6$ matrix. after some simple calculations, we give the explicit expression  
of the matrix below :\\~~~~~\\~~\\

$f= $\bordermatrix{~ & X^1_0 & X^2_0 &  X^1_1 & X^2_1 &  P^1_0 & P^2_0 \cr
                  X^1_0  & 0& -2(B^{-1}M)_{12}  & 0 & 0 & 0 & 0\cr
                  X^2_0  & 2(B^{-1}M)_{12}  & 0 & 0 &0&0 & 0 \cr
                 X^1_1 & 0 & 0 & 0&2 m^2 B_{12}&-1 & 0\cr
                  X^2_1 & 0 & 0 & -2 m^2 B_{12}&0&0 & -1 \cr
                 p^1_0 & 0 & 0 &1&0& 0& 0\cr
                  p^2_0& 0 & 0 & 0 &1&0& 0 \cr}\\~~~~~\\~~\\
 \\~~~\\~~~\\
$ f $ is not singular provided $ B_{\mu\nu} $ no vanishing , hence the inverse of this matrix exists :\\~~\\
$f^{-1}= $\bordermatrix{~ & X^1_0 & X^2_0 &  X^1_1 & X^2_1 &  P^1_0 & P^2_0 \cr
                  X^1_0  & 0& \frac{(BM^{-1})_{12}}{2}  & 0&0&0 & 0\cr
                  X^2_0  & -\frac{(BM^{-1})_{12}}{2}  & 0 & 0 &0&0 & 0 \cr
                  X^1_1 & 0 & 0& 0&0&1 & 0\cr
                  X^2_1 &0  & 0 & 0&0&0 & 1 \cr
                  p^1_0  & 0 & 0 &-1&0& 0& 2m^2 B_{12}\cr
                  p^2_0 & 0 & 0 & 0 & -1 & -2m^2 B_{12} & 0 \cr}\\~~~\\~~\\
                  From the above matrix $f^{-1} $ ,we can read the following commutators,

                  \begin{align}
                  [X^\mu_0,X^\nu_0]=&2\pi\alpha'(BM^{-1})^{\mu\nu}\\
                 [P^\mu_0,P^\nu_0]=&\frac{m^2 B^{\mu\nu}}{2\pi\alpha'}\\
                 [X^{\mu}_1,P^\nu_0]=&2\pi\alpha'\delta^{\mu\nu}                
                  \end{align}
                  Here we recovered the coefficient $2\pi\alpha' $ explicitly.
\section{Conclusion}
As we see from Eqs. (29) and (30), for $ m\neq 0 $, the momentum fields as well as the
coordinate fields are noncommutative at the end-points. This is due to dependence of momentum fields to $ m $, so the whole phase space becomes noncommutative to lead to the so called totally noncommutativity.
This situation is even more difficult than the case of noncommutative spacetime where
we were allowed to view momentum operators as differential operators. Equivalently, we
were allowed to use momentum space representation because the momentum space was
commutative. However, in totally noncommutative phase space, neither coordinate space
representation nor momentum space representations are allowed.
So in order to study physics on totally noncommutative phase space, we may propose
to use the generalisation of phase space quantisation. However, it is difficult to construct a quantum field
theory on totally noncommutative phase space. This is because we normally view fields
as functions on either spacetime or momentum space. However, we are now forced to
use phase space functions. We might need to first construct quantum field theory in
usual phase space. Then the next step will by easy: as the fields are already functions
of phase space, we may use star-product to give the details of totally noncommutative
phase space. Alternatively, we may first try to understand some physics by studying
quantum mechanics in totally noncommutative phase space. This might provide some
useful insights into constructing quantum field theory in totally noncommutative phase
space.\\


\begin{thebibliography}{1}
  \bibitem{impj}  E. Witten {\em "Noncommutative Geometry And String Field Theory" }Nucl. Phys. B 268
(1986) 253.
  \bibitem{norman}Seiberg, Nathan et al. {\em "String Theory and Noncommutative Geometry"}   JHEP 9909 (1999) 032 hep-th/9908142 IASSNS-HEP-99-74 .
  \bibitem{} A.Connes ,M.R.Douglas, and.Shwarz {\em "noncommutative geometry matrix theory : compactification on tori"} JHEP 02 (1998) 003, [hep-th/9711162].
\bibitem{}F.Ardalan, H. Arfaei, and M. M. Sheikh-Jabbari {\em "Dirac Quantization of Open Strings and Noncommutativity in Branes"}
Nuclear Physics B 576 (1-3), 578-596.
\bibitem{}Won Tae Kim and John J. Oh {\em "Noncommutative open string from Dirac quantization"} Mod.Phys.Lett. A15 (2000) 1597-1604.
\bibitem{}] Ralph Blumenhagen {\em "A Course on Noncommutative Geometry in String "} Theory Max-Planck-Institut fur Physik, Fohringer Ring 6, 80805 Munchen, arXiv:1403.4805v2 [hep-th] 1 Apr 2014.
\bibitem{}Louise Dolan, Chiara R. Nappi{\em "Strings and Noncommutativity"} ,[ hep-th/0302122].
\bibitem{} Chong-Sun Chu{\em "noncommutative geometry from strings"} , [hep-th/0502167].
\bibitem{}C. Hofman and E. Verlinde {\em "U-Duality of Born-Infeld on the Noncommutative Two-Torus"} JHEP 12 (1998) 010, [hep-th/9810116].
\bibitem{}C.-S. Chu and P.-M. Ho {\em "Noncommutative Open String and D-brane"} Nucl. Phys. B550 (1999) 151, [hep-th/9812219].
\bibitem{}Xiao-Jun Wang {\em "Strings in Noncommutative Spacetime"} ,[hep-th/0503111].
\bibitem{}Chong-Sun Chu, and Pei-Ming Ho {\em "Constrained Quantization of Open String in
Background B Field and Noncommutative D-brane
"}NEIP-99-011[hep-th/9906192].
\bibitem{}Farhad Ardalan, Hessamaddin Arfaei and Mohammad M. Sheikh-Jabbari{\em "Noncommutative geometry from strings and branes "} JHEP02(1999)016 , http://iopscience.iop.org/1126-6708/1999/02/016.
\bibitem{}Nelson R. F. Braga and Cresus F. L. Godinho{\em "Symplectic Quantization of Open Strings and Noncommutativity in Branes"} Phys.Rev.D65:085030,2002.
\bibitem{}L. Faddeev , R. Jackiw , Phys.Rev.Lett.60 (1988) 1692.
\bibitem{}Pichet Vanichchapongjaroen {\em " D-Branes and Noncommutative Geometry
in String Theory"}
\bibitem{}A.Shirzad A. Bakhshi and Y. Koohsarian{\em "Symplectic Quantization of Massive Bosonic String in background B-field"} Mod. Phys. Lett. A, Vol. 27, No 13 (2012) 1250073.
\bibitem{}L.D. Faddeev,R.Jackiw,Phys.Rev.Lett.60(1988)1691.
\bibitem{}N. Mansour, E. Y.Diaf and M.B.Sedra : EJTP 14 21, 34 (2018)

  \end{thebibliography}
\end{document}